\documentstyle[pra,aps,epsf,twocolumn,amssymb]{revtex}
\begin{document}

\title{Truncations of random unitary matrices}

\author{ Karol  \.Zyczkowski$^{1,2}$ and Hans-J{\"u}rgen Sommers$^3$}

\address{
  $^{1}$Instytut Fizyki Mariana Smoluchowskiego, Uniwersytet
Jagiello\'nski, \\
ul. Reymonta 4,  30-059 Krak\'ow, Poland\\
   $^{2}$Centrum Fizyki Teoretycznej PAN  \\
      Al. Lotnik{\'o}w 32/46, 02-668 Warszawa, Poland \\
  $^3$Fachbereich 7 Physik, Gesamthochschule Essen, D-45117 Essen,
Germany}

\date{\today}
\maketitle
\begin{abstract}
We analyze properties of non-hermitian matrices of size $M$
constructed as  square submatrices  of unitary (orthogonal) random
matrices of size
$N>M$, distributed according to the Haar measure.
In this way we define ensembles of random matrices and study the
statistical properties of the spectrum located inside the unit circle.
In the limit of large matrices, this ensemble is
characterized by the ratio $M/N$. For the truncated CUE we derive
analytically the joint density of eigenvalues
and all correlation functions. In the strongly non unitary case universal
Ginibre behaviour is found.
 For $N-M$ fixed and $N$ to $\infty$ the universal
resonance-width distribution with $N-M$ open channels is recovered.
\end{abstract}

\section{Introduction}

Random unitary matrices may be applied to describe
 chaotic scattering \cite{bs88}, conductance in mesoscopic
systems\cite{Be97} or statistical properties of periodically driven
quantum systems  (see \cite{H91} and references therein).
 They can be defined by circular ensembles of unitary
matrices introduced by Dyson \cite{dys63}. He defined circular
orthogonal, unitary or symplectic ensembles (COE, CUE and CSE), which display
different transformation properties \cite{mehta}.  For these ensembles
the  distribution of matrix elements and their correlations are known
\cite{pm83,mps85,m90}.

In the present paper we discuss properties of non-hermitian
matrices defined as square submatrices of size $M$ of
unitary (orthogonal) matrices of size $N$, where $N>M$.
 These matrices may be
considered as unitary (orthogonal) matrices with $N-M$ bottom rows and
$N-M$ last columns truncated.
Let  $U_{[N,M]}$ denote such a $M\times M$ matrix obtained from a
unitary matrix, while $O_{[N,M]}$ is obtained by truncating an
orthogonal matrix. The truncated matrices
are non-unitary by construction, and their eigenvalues are located
inside the unit circle.

Motivation for such a study stems from the problems of
chaotic scattering. Consider
a mesoscopic device coupled to two leads, each of which supports $N/2$
open channels. The process of scattering can be described by a unitary
$S$-matrix of size $N$.  In the diffusive regime the scattering
matrix pertains to an appropriate circular ensemble \cite{Be97}. The
reflection (transmission) matrix of size $M=N/2$
may be just considered as a truncation of the unitary $S$-matrix.
The random matrix approach to resonances in chaotic scattering was
 recently presented in \cite{FS97}. In particular, the distribution of
width of resonances in the presence of $L$ open channels was derived
in the weakly non hermitian limit for broken time reversal symmetry.

In  recent papers \cite{Kol99a,Kol99} the authors introduce  $N\times N$
unitary matrices enlarged in an asymmetric way to the size $(N+L)\times
(N+L)$ by adding
$L$ {\sl upper} rows and $L$ {\sl last}
 columns with all elements equal to zero.
 These matrices are used to describe the chaotic
scattering in a
1D model of crystal electron in ac and dc fields.
 It is easy to see, that the spectrum of such an {\sl
enlarged} matrix consists of $2L$ zeros and $M=N-L$ complex
eigenvalues of the {\sl truncated} matrix $U_{[N,N-L]}$. Our results are
therefore directly applicable to the problems analyzed in that papers.

Related problems arise by analyzing the time evolution
of periodically perturbed systems. The model of the kicked rotator
with absorbing boundaries was studied in \cite{BGS91,CMS97,CMS99}.
In this case the presence of the absorbing boundaries corresponds to
the truncation of the infinite evolution matrix.
 Another line of
 research is related to the Frobenius-Perron operators
describing the evolution of densities under a classical map. If the map
is area preserving, the FP-operator is represented by an infinite
unitary matrix, which for practical purposes is truncated to a finite
size.  Properties of the spectra of such truncated matrices
  have recently attracted a considerable attention \cite{Fi99,HW99}.

This paper is organized as follows. In section II
we analyze  truncations of orthogonal matrices and show a
geometric interpretation of this problem. We derive the
probability distributions of the radii of points uniformly
covering a given hypersphere and projected into a smaller space.
Section III is devoted to truncations of random unitary matrices. We
demonstrate a
link between the distributions studied and the eigenvector statistics.
In section IV we present numerical results concerning the
distribution of the complex eigenvalues
of the truncated matrices . We show to what extent the ratio $\mu =M/N$
determines the
properties of the truncated matrix.
In section V we analytically derive the joint density of eigenvalues for
truncated matrices of CUE. From this a kernel is derived which determines all  
correlation functions. In particular the averaged density of eigenvalues is  
obtained for arbitrary dimensions and truncations.
In the strongly nonunitary limit $ M \to \infty,\ N/M\ fixed$ the  
correlations of the Ginibre ensemble \cite{Ginib} are obtained rescaled by  
the local mean level distance, which are thus revealed as universal.
The weakly non unitary
limit $N-M$ fixed and $M$ to $\infty$ recovers the universal distribution of  
resonance widths in the weakly non-hermitian case  for broken time reversal  
symmetry
\cite{FS97} and the corresponding correlations \cite{FK99}.
The truncations of symmetric matrices of COE are briefly discussed in
section VI. Convolution properties of the derived distributions are
presented in the Appendix.

\section{Submatrices of random orthogonal matrices}

Let us start the discussion considering  a simple geometric
exercise. Random points cover uniformly a hypersphere $S^{N-1}$ of
radius $1$ embedded in $R^N$.
After an orthogonal projection into $R^M$, where $M<N$,
they are localized inside  the hypersphere
$S^{M-1}$ or at its surface.
What is the radial probability distribution $P_{N,M}(t)$,
where
$t$ denotes the distance of a projected point from the origin?

It is helpful to analyze first the  most intuitive case $N=3,M=2$.
The surface element of the sphere $S^2$ in
spherical coordinates reads
$d\Omega_2=\sin\theta d\theta d\phi$. The orthogonal projection maps the
points of the sphere into a plane. Their distance from
the origin is $t=\sin\theta$, which allows us to find the required
distribution
\begin{equation}
     P_{3,2}(t)= \frac{t}{2\sqrt{1-t^2}}.
\label{po32}
\end{equation}
Analogously we get $P_{3,1}(t)=1$ for $t\in [0,1]$ and
$P_{2,1}(t)=1/(2\pi\sqrt{1-t^2})$.

The general formula for $P_{N,M}(t)$ may be obtained in a similar
way from the element of the hypersphere $S^{N-1}$
\begin{equation}
d\Omega_{N-1}=d\varphi\Pi_{k=1}^{N-2}  \sin^k\theta_k d\theta_k.
\label{domega}
\end{equation}
Integrating out $N-M$ variables we obtain
\begin{equation}
     P^{o}_{N,M}(t)= c_{N,M}\ \ t^{M-1}(1-t^2)^{(N-M-2)/2},
\label{ponm}
\end{equation}
where the normalization constant can be expressed by the Euler beta
function $B(x,y)$ \cite{grad}
\begin{equation}
 c_{N,M}=\frac{2}{B({M\over 2}, {N-M \over 2})} =
 { 2\Gamma({N \over 2}) \over
  \Gamma({M \over 2}) \Gamma({N-M \over 2}) }.
\label{cnm}
\end{equation}
Convolution relations between the distributions $P^o_{N,M}(t)$ are
demonstrated in the Appendix.

Consider an orthogonal matrix $O$ of size $N$. Its first column
can be interpreted as a vector $x_k=O_{k1}$ of coordinates
determining a point on the hypersphere $S^{N-1}$.
 Let us call by $O_{[N,M]}$ the upper left submatrix of $O$
of size $M<N$.
The total length of the vector represented by the first column of
$O_{[N,M]}$ and given by $t=\sqrt{ \sum_{k=1}^M x_k^2}$ is just
equal to the defined above distance of a point projected
from the hypersphere $S^{N-1}$ into the interior of $S^{M-1}$ from the
origin. If $O$ are distributed uniformly with respect to the
Haar measure on $O(N)$, than the points $x$ cover uniformly the
hypersphere. The distributions $P^o_{N,M}(t)$ are then given by Eq.
(\ref{ponm}).

Figure 1.a shows these distributions for $N=16$ and $M=1,2,4,8$ and
$15$.
With increasing $M$ the probability distribution is shifted to the
right. For $M=N$ the matrix remains unitary and
$P_{N,N}(t)=\delta(t-1)$. Let us  now
consider an ensemble $O_{[N, \mu N]}$ by increasing the dimension $N$ and
keeping the ratio $\mu =M/N$ fixed, where $\mu <1$ .
Straightforward integration allows us to compute for this ensemble
the expectation value of $t$
\begin{equation}
 \langle t\rangle_{N,\mu N}=
 { \Gamma({N \over 2}) \Gamma({\mu N \over 2}+{1 \over 2})
  \over
   \Gamma({N \over 2}+{1 \over 2})
  \Gamma({\mu N \over 2})},
\label{texp}
\end{equation}
which in the limit $N\to\infty$ tends to $\sqrt{\mu }$.
The second moment reads
 $\langle t^2 \rangle_{N,\mu N}=\mu$, thus the variance tends to zero in
the limit of large matrices. This result is quite intuitive in
view of the central limit theorem.

\vskip -1.4cm
\begin{figure}
\hspace*{-1.6cm}
\vspace*{-1.2cm}
\epsfxsize=10.5cm
\epsfbox{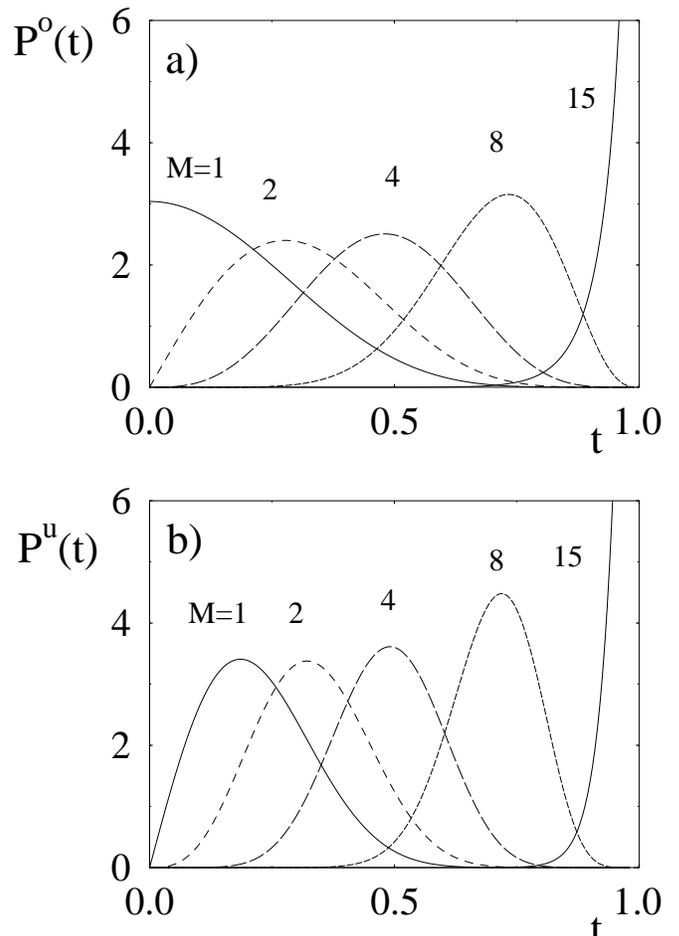}
\caption{Probability distribution of the radii $t$ after projection
from $R^{16}$ to $R^M$ with  $M=1,2,4,8,15$. The variable $t^2$
is equal to the sum of  $M$ squared elements of a random orthogonal
matrix distributed according to the Haar measure on $O(16)$ (a).  Analogous
distributions $P^u(t)$ obtained from random unitary matrices pertaining
to $U(16)$, (b).}
\label{fig1}
\end{figure}

\section{Submatrices of random unitary  matrices}

 Let $U_{[N,M]}$ denote the $M\times M$ matrix obtained by truncation
of the $N \times N$ unitary matrix $U$. In a similar way we define
$t=\sqrt{ \sum_{k=1}^M |U_{k1}|^2}$.
To find in this case the distribution $P^u_{N,M}(t)$ it is
useful to represent a unitary matrix $U(N)$ as a product
(the normal product $(\times)$
or the twisted product $(\ltimes)$) of the hyperspheres \cite{Boya}
\begin{equation}
      U(N) \sim S^1 \times  S^3 \ltimes \cdots \ltimes S^{2N-3} \ltimes
S^{2N-1}.
\label{udec}
\end{equation}
Truncation of the dimension of a unitary matrix by one corresponds
to the projection from $S^{2N-1}$ to $S^{2N-3}$, which is equivalent to
the truncation of the matrix $O(2N)$ by two. The same argument works for
any size $M$ of the truncated matrix. Therefore
$P^u_{N,M}(t)=P^o_{2N,2M}(t)$ and
\begin{equation}
     P^{u}_{N,M}(t)= c_{2N,2M} \ \ t^{2M-1}(1-t^2)^{N-M-1},
\label{punm}
\end{equation}
with the normalization constants given by (\ref{cnm}). Some of
these  distributions  for $N=16$ are plotted in Fig 1b.
Expectation values $\langle t\rangle$ are asymptotically the same for
the both ensembles, while the variance is smaller
for the ensemble of truncated unitary matrices $U_{[N,M]}$.

For a fixed value of $N$ we defined the ensembles of truncated
matrices for each integer value of $M\in [1,N]$. However, to study
the evolution of a spectrum of a given matrix
it is convenient to define an ensemble
depending on a continuous parameter. This can be achieved in several
different ways. For example,
one may multiply the last column and the last vector of the matrix
$U_{[N,M]}$ by a parameter $p \in [0,1]$, which mimics a continuous
transition from $M$ to $M-1$ \cite{HW99}.

Taking $M=1$ the variable $t$ is just the absolute value of the first
element of a matrix $|U_{11}|$. It is known \cite{mehta,girko} that a
unitary matrix
of eigenvectors of a CUE matrix is distributed according to the Haar
measure on $U(N)$, while the orthogonal matrix of eigenvectors of a
matrix
typical of COE is distributed according to the Haar measure on $O(N)$.
To establish a link with the eigenvector statistics let us set
$M=1$ and consider the distributions $P^o_{N,1}(t)$ and
$P^u_{N,1}(t)=P^o_{2N,2}(t)$. Putting  $y=t^2$ and changing the variable
we arrive at the known formulae
\begin{equation}
 P^o_N(y) =
 { \Gamma({N \over 2}) \over \Gamma({N -1 \over 2}) }
 {(1-y)^{(N-3)/2} \over \sqrt{\pi y} },
\label{pony}
\end{equation}
 and
\begin{equation}
 P^u_N(y) = (N-1)(1-y)^{N-2},
\label{puny}
\end{equation}
which describe the eigenvector statistics for the orthogonal and
the unitary ensemble \cite{KMH88,HZ90}. In the limit $N\to \infty$ they
converge
to the $\chi^2_{\nu}$ distributions with the number of degrees of
freedom $\nu$ equal to $1$ and $2$, respectively. The former case
is often known in the literature as the Porter-Thomas distribution.

\section{Distribution of eigenvalues }

Consider spectra of the truncated orthogonal matrices  $O_{[N,M]}$ and
truncated unitary matrices $U_{[N,M]}$. In both cases there exist $M$
complex eigenvalues $z_j=r_j\exp(i\phi_j)$ localized inside (or at) the
unit circle. This is due to the fact that the norm of the
truncation is smaller than or equal to the norm
of the initial matrix.
For the truncations of random matrices of CUE
there exist an rotational symmetry, $U\to U\exp(i\alpha)$.
Therefore $P(\phi)=const$, so
 we will study the radial distribution $P(r)$.
Sometimes it is convenient to write $r=e^{(-\gamma/2)}$
and to study the distribution $P(\gamma)$ of the "level widths"
 $\gamma$ \cite{BGS91}.
For any fixed $N$ the limiting cases are known: for $M=1$
the eigenvalues are trivial, $r=t$, so for both ensembles
$P_{N,1}(r)=P_{N,1}(t)$. For $M=N$ the matrix is unitary and thus
$P_{N,N}(r)=\delta(r-1)$ or, in other variables,
$P_{N,N}(\gamma)=\delta(\gamma)$.

\vskip -0.2cm
\begin{figure}
\hspace*{-0.1cm}
\vspace*{-0.1cm}
\epsfxsize=7.8cm
\epsfbox{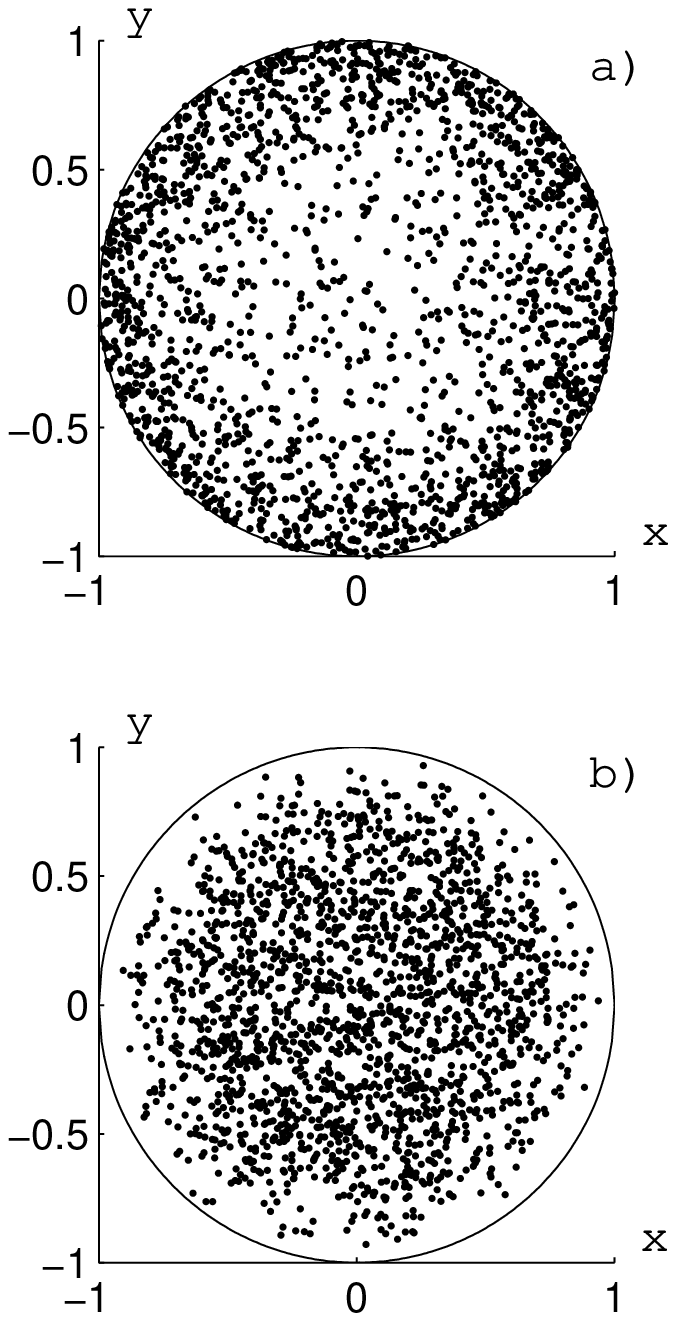}
\caption{Complex eigenvalues of random matrices pertaining to
 the ensembles $U_{[5,4]}$ (a) and $U_{[5,2]}$ (b).}
\label{fig2}
\end{figure}

Figure 2 presents 2000 eigenvales of the matrices truncated out of CUE
matrices of size $5$. For $M=4$ there exist several eigenvalues
close to the unit circle, while for stronger truncation ($M=2$)
the eigenvalues are clustered closer to the origin.

In the simplest interesting case, $N=3$ and $M=2$
  the data for the truncations of unitary matrices conform
to the distribution $P^u_{3,2}(r)=r+2r^3$.  For comparison with the
results of \cite{BGS91} we present the numerical data
as the distribution $P(\gamma)$. The above distribution,
derived in the following section,  corresponds to the
biexponential decay
\begin{equation}
 P^u_{3,2}(\gamma) = {1 \over 2} \exp(- \gamma) + \exp(-2\gamma).
\label{pgam}
\end{equation}
represented by a solid line in Fig. 3a. Numerical data obtained for
the ensemble $U_{[5,4]}$, shown in Fig. 3b, are compared
with the distribution
$ P^u_{5,4}(\gamma) = {1 \over 4} \exp(-\gamma)
+ {1 \over 2} \exp(-2 \gamma) +
  {3 \over 4} \exp(-3 \gamma) + \exp(-4\gamma )$,
which corresponds to $P^u_{5,4}(r)$ discussed below.

\vskip -1.6cm
\begin{figure}
\hspace*{-2.0cm}
\vspace*{-1.2cm}
\epsfxsize=10.0cm
\epsfbox{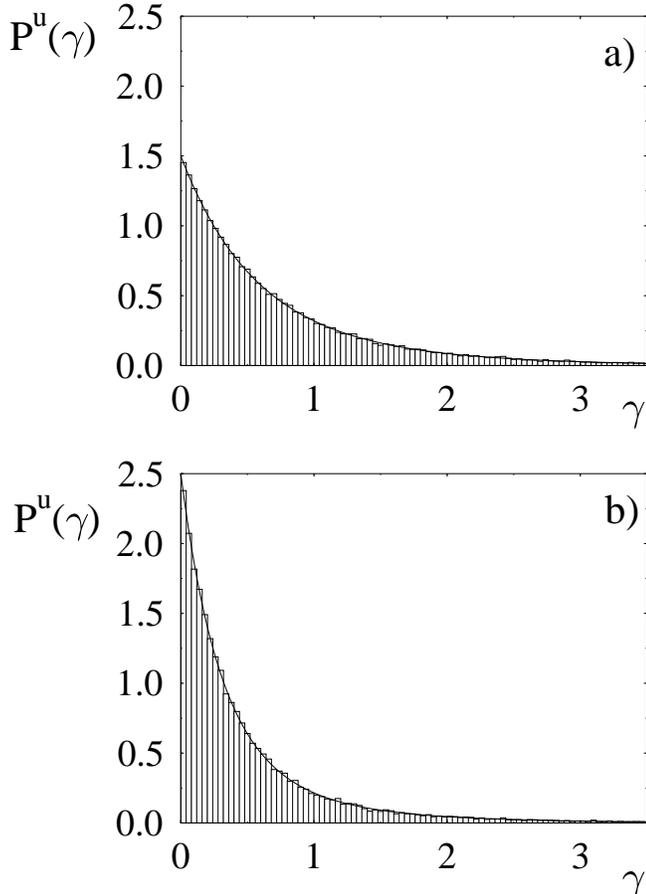}
\caption{Distribution of decay times $P(\gamma)$
for random matrices pertaining to
 the ensembles $U_{[3,2]}$ (a) and $U_{[5,4]}$ (b).}
\label{fig3}
\end{figure}

 Distributions $P(r)$ for both ensembles obtained with
$N=16$ and some intermediate values of $M$ are presented in Fig. 4.
The histograms are performed out of $10^4$ random unitary (orthogonal)
matrices constructed according to the algorithm given in \cite{PZK98}.
The statistics
obtained do not depend on which columns and rows of the initially
unitary (orthogonal) matrix are removed during the truncation.
This is due to the fact that the Haar measure on
the unitary (or orthogonal) group is invariant with respect to
multiplication by the permutation matrices, which change the order
of the columns and vectors.

 With increase of $M$ the  distribution
$P(r)$ extends to the larger values of $r$.
In contrast with the distributions $P(t)$,
for any $M$ there exists a non-zero probability of finding small values
of $r$. For small values $r$ the distribution
$P^u(r)$ grows linearly with $r$.
This is a purely geometric factor (we analyze the distribution at
the complex plane), which corresponds to the uniform density of
eigenvalues close to the origin.

\vskip -1.4cm
\begin{figure}
\hspace*{-1.6cm}
\vspace*{-1.3cm}
\epsfxsize=10.5cm
\epsfbox{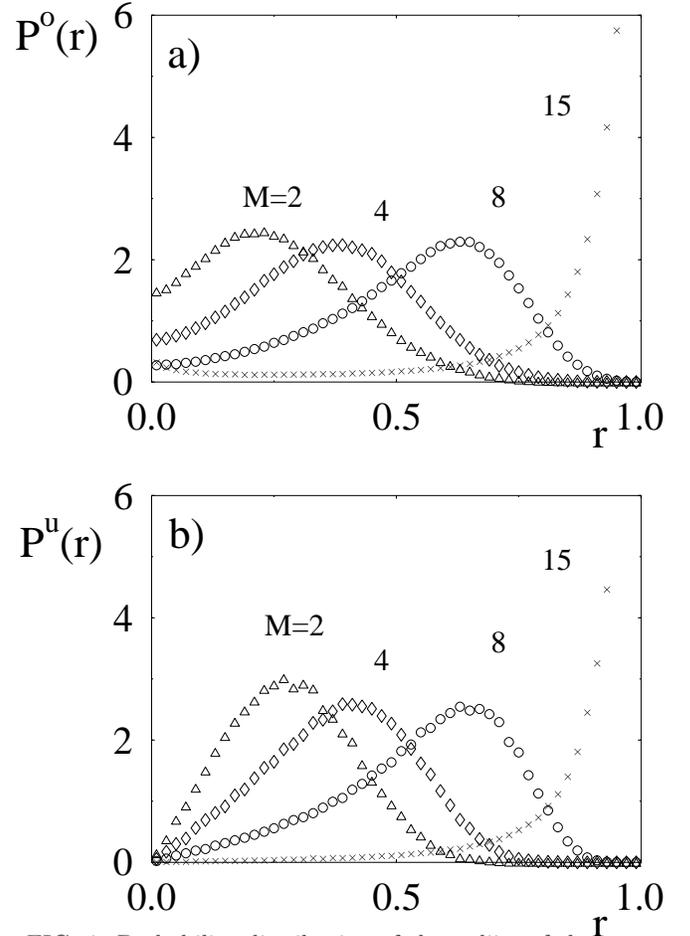}
\caption{Probability distribution of the radii $r$ of the complex
eigenvalues of the $M\times M$ sub-blocks of random matrices of $O(16)$
(a). Analogous  distributions $P^u(r)$ obtained
 from random unitary matrices pertaining
to $U(16)$, (b).}
\label{fig4}
\end{figure}

The data collected for large matrices reveal a scaling behavior:
the distribution $P_{N,M}(r)$ depends only on the ratio $\mu=M/N$.
Figure 5 shows the distributions $P^u_{N,N/2}(r)$ and
 $P^u_{N,N/4}(r)$ obtained from ensembles of random unitary
matrices of different sizes. The larger value of $N$, the sharper
is the cut-off of the probability at the critical radius
$r_{\mu}=\sqrt{\mu}$.
In analogy to the properties of the Ginibre
ensemble one  expects an infinitely sharp edge of the distribution in
the limit $N\to \infty$.
In the case of large matrices
the spectrum covers the entire circle of radius $r_{\mu}$, while the
density is largest close to the rim.

\vskip -1.6cm
\begin{figure}
\hspace*{-1.7cm}
\vspace*{-1.1cm}
\epsfxsize=10.5cm
\epsfbox{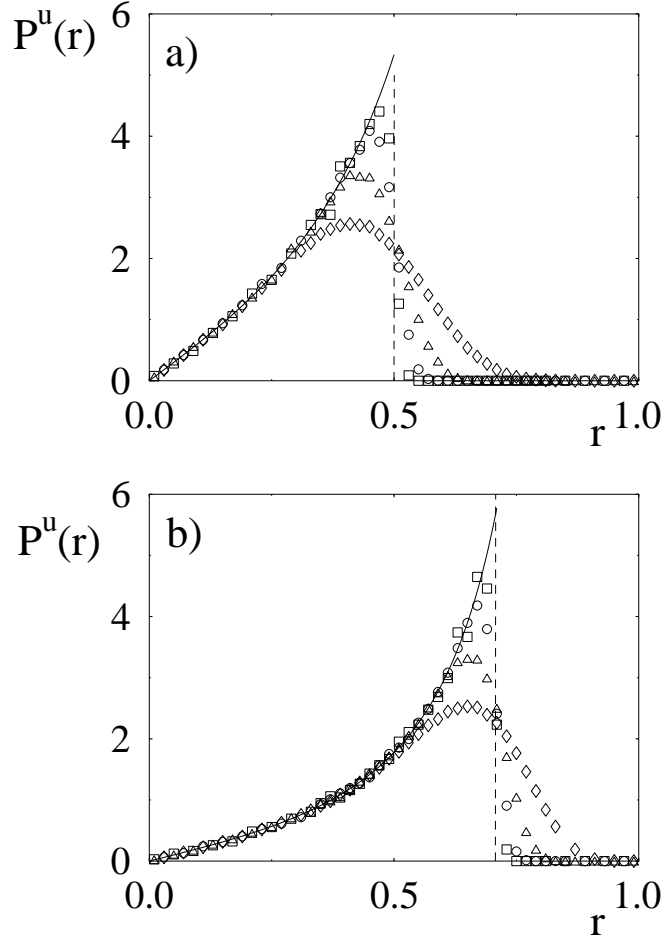}
\caption{Probability distribution $P^u(r)$ of radii of eigenvalues
of matrices $U_{[N,M]}$
constructed from unitary matrices of size $N=16(\diamond ),$ $64(\triangle )$,
$ 256 (\circ ) $ and
$1024 (\square )$ with $M=N/4$ (a), and $M=N/2$ (b).
Dashed lines represent the asymptotic cut-off at $r=r_{\mu}$ and the
solid line denotes the distribution (\ref{prmu}). }
\label{fig5}
\end{figure}

 For $\mu << 1 $ (and $N$ large)
the radial distribution may be approximated by a linear function
 $P^u_{\mu}(r)\sim 2r/{\mu}$ with a cut-off at $r_{\mu}$.
 This property is characteristic of the Ginibre
ensemble \cite{Ginib}, constructed of non-Hermitian random matrices with
no correlations between their elements. It is thus
intuitive to expect, that for large $N$ the constraints
stemming from the unitarity of $U(N)$ do not induce very strong
correlations between elements of a much smaller matrix of size $M$.

Eigenvalues of several truncations of random orthogonal
matrices are shown in Fig. 6. Since the truncated matrix is real
the eigenvalues are real   or appear in complex conjugate
pairs, $r e^{i\phi}$, $r e^{-i\phi}$. Therefore these spectra
exhibit the symmetry along the real line.
 Observe a clustering of eigenvalues along this line.
The fraction of real eigenvalues equals approximatelly
 $0.65$, $0.38$ and $0.68$ for the ensembles
$O_{[3,2]}$, $O_{[5,4]}$ and $O_{[5,2]}$, respectively.
This fact explains a positive probability $P^o(r)$
for $r=0$ visible in Fig. 4a.

\vskip -0.2cm
\begin{figure}
\hspace*{-0.1cm}
\vspace*{-0.0cm}
\epsfxsize=7.8cm
\epsfbox{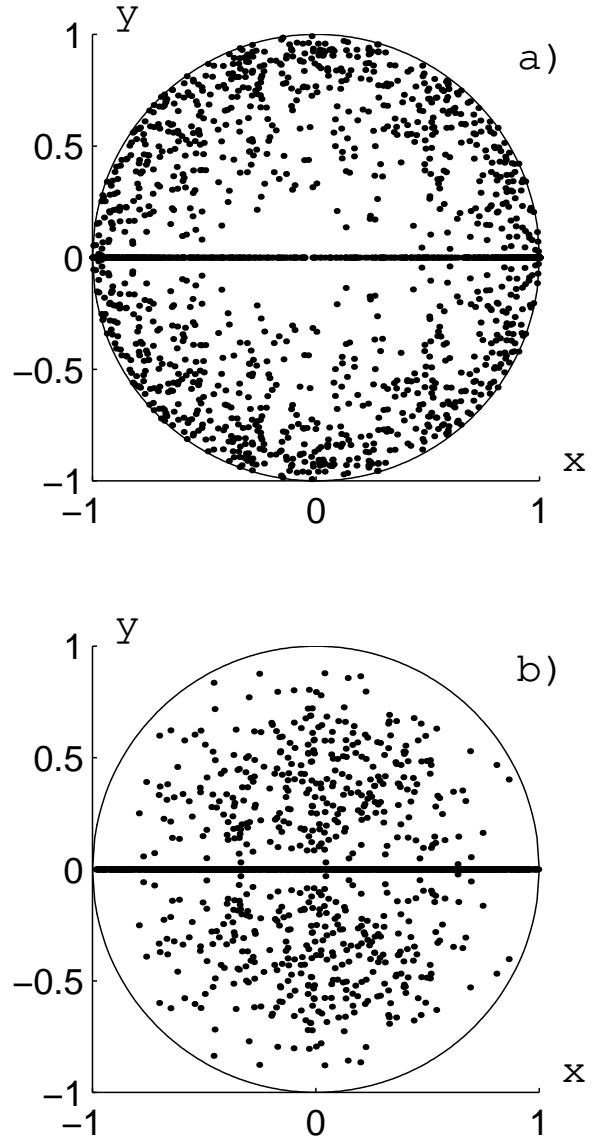}
\caption{As in Fig.2 for truncations of random orthogonal matrices
 a) $O_{[5,4]}$ and b) $O_{[5,2]}$. Note clustering
of eigenvalues at the real axis.}
\label{fig6}
\end{figure}

If the ratio $\mu$ is kept constant,
the relative number of real eigenvalues
decreases with the matrix size.
Similar effect is known in the theory of random polynomials.
Kac considered the random polynomials of order $M$
with  real coefficients, being independent random variables
drawn according to the normal distribution. He showed
\cite{Kac} that the fraction of real roots
decreases as $(\ln M)/M$. Our problem is not exactly the same
since the real coefficients of the secular polynomial
of the truncated random matrix are not Gaussian,
nor independent random variables.
In spite of this fact, our numerical results suggest
that the fraction of the real eigenvalues of truncations
of orthogonal matrices $O_{[2M,M]}$ decreases as $(\ln(M)/M)$.
Recent discussion
of properties of random polynomials and their
applications to quantum chaos may be found in \cite{BBL96}.
The issue of clustering of zeros of random polynomials along a
given curve and its relation to the time reversal symmetry is
discussed in \cite{BKZ97}.

In the limit of large matrices the relative strength
of the clustering of the complex eigenvalues along  the real
axis decreases and the distribution  $P(\phi)$
becomes uniform. Moreover, the radial distribution
$P^o(r)$ becomes close to the distribution (\ref{prmu}) derived for
truncations of unitary matrices. Although for $N=16$
the differences between the distributions $P^o_{\mu}(r)$ and
$P^u_{\mu}(r)$
are significant, especially for small values of $r$,
for large $N$ the data for both ensembles seem to converge.

\section{Analytical results for truncations of CUE }

Now we derive analytical results for the truncated circular unitary ensemble.
Let $$ U = \left(\matrix{A & B \cr
                         C & D \cr}
           \right)$$
be an $ N \times N $ matrix from this ensemble and $A$ a subunitary $ M
\times M $ matrix.
The joint density of elements of $U$ can be written as

\begin{eqnarray}
 P(U) \propto \delta (A^{\dagger}A+C^{\dagger}C-1)
 \delta (A^{\dagger}B+C^{\dagger}D)
\times \nonumber \\
 \delta (B^{\dagger}B+D^{\dagger}D-1)
\label{G1}
\end{eqnarray}
with appropriate matrix $\delta-$functions. Integrating out $B$ and $D$
we obtain as joint density of elements of $A$

\begin{equation}
 P(A) \propto  \int dC \delta (A^{\dagger}A+C^{\dagger}C-1)
\label{G2}
\end{equation}
with a $2M(N-M)$ dimensional integration over the complex parameters $C$.
The matrix $A$ may be brought to upper triangular form by a unitary
transformation $T$ : $A = T(z+ \Delta)T^{-1}$, where $z$ is a diagonal
matrix consisting of the complex eigenvalues of $A$ and $\Delta$ is strictly
upper triangular ( Schur decomposition).
The transformation can be made unique restricting $T$ to a certain cosetspace.
The Jacobian of this transformation
is given by the square of the Vandermonde determinant \cite{FKS97}
  $$ \vert V \vert^2 = \prod_{i<j} \vert z_i -z_j\vert^2 $$
such that after integrating out the unitary transformations $T$ the joint
density of eigenvalues is given by

\begin{equation}
 P(z) \propto \vert V \vert^2 \int dC \int d\Delta \delta ((z^{\dagger} +
\Delta^{\dagger})(z + \Delta) +C^{\dagger}C-1).
\label{G3}
\end{equation}
In the following we first integrate out the $M(M-1)/2$ complex parameters
$\Delta_{ij}$ and then the $M(N-M)$ complex parameters $C$ written as complex  
vectors $C_i$.
For $i<j$ we have the hierarchical equations
\begin{equation}
z_i^* \Delta_{ij} + C_i^{\dagger}C_j + \sum_{k<i}\Delta_{ki}^*\Delta_{kj} =0.    
\label{G4}
\end{equation}
Integration over $\Delta$ yields the Jacobian $\prod_{i<j}\vert z_i\vert^{-2}$ 
and a product of $\delta-$functions
\begin{equation}
\prod_i \delta(\vert z_i\vert^{ 2} + C_i^{\dagger} X_i C_i -1).
\label{G5}
\end{equation}
where $X_i$ denotes an $N-M \times N-M$ matrix defined by
the quadratic form $C_i^{\dagger}X_iC_i$ which is given by
 $$C_i^{\dagger}C_i + \sum_{k<i}\Delta_{ki}^*\Delta_{ki}$$
containing the solution $\Delta_{ki}$ of equ.(\ref{G4})
and depending otherwise on $C_k$ only for $k<i$. The integration over $C_i$  
can now be done successively starting from $C_M$ and yields the factor
$$ \prod_{i=1}^M(1-\vert z_i\vert^{ 2})^{N-M-1}\Theta (1-\vert
z_i\vert^{ 2})/ \det(X_i)$$
 where $\Theta(.)$ denotes the Heaviside step function,
 $\Theta(x)=1$ for $x>0$ and zero otherwise.

 For $\det(X_i)$ we can derive from the equations (\ref{G4}) for
$\Delta$ and using implicitly the $\delta-$functions (\ref{G5}) the recursive 
relation $$\det(X_{i+1}) = \det(X_{i })/\vert z_i\vert^{ 2}$$
with $\det(X_1)=1 $.  Thus the previous Jacobian $\prod_{i<j}\vert z_i\vert^{-2}$ 
will be compensated and the final very simple and important result is
\begin{equation}
P(z) \propto \prod_{i<j}^{1.. M}\vert z_i-z_j \vert^2 \prod_{i=1}^M (1-\vert  
z_i\vert^{ 2})^{N-M-1}\Theta(1-\vert z_i\vert^{ 2})
\label{G6}
\end{equation}
This result is completely analogous to the Ginibre  ensemble and we know
immediately all correlation functions by the method of orthogonal
polynomials
\cite{mehta}.
Here the powers $z^{n-1}$ are already orthogonal. An equivalent
method is to
consider the joint density $P(z)$ as the absolute square of a Slater
determinant of normalized wave functions
$$ \phi_n(z) = z^{n-1} w(\vert z\vert^2)/\sqrt{N_n}$$
with $$ w(x)=(1-x)^{(N-M-1)/2}\Theta(1-x),$$
where $N_n$ stands for an normalization factor.

The kernel, which determines all correlation functions is
\cite{mehta,FKS97} $$
K(z_1,z_2^*) = \sum_{n=1}^M (z_1z_2^*)^{n-1} w(\vert z_1\vert^2)
w(\vert z_2\vert^2) /N_n.$$ For example the cluster function is given
by $ Y(z_1,z_2)= \vert K(z_1,z_2^*)\vert^2$ and the averaged density
of eigenvalues z normalized to $1$ is given by

\begin{equation} \rho (z) =
K(z,z^*)/M={1\over M} \sum_{n=1}^M \vert z \vert^{ 2n-2}w^2(\vert z
\vert^2)/N_n.
\label{G7}
\end{equation}
The normalisation factor
$N_n$ is easily calculated as
 $$ N_n = \pi (n-1)!(N-M-1)!/(N-M+n-1)!$$
For example with $ r^2=\vert z\vert^2$ we
obtain for the distribution of r with $M=N-1$
$$ P(r)= {2\over M}(r+2r^3+3r^5+...+Mr^{2M-1})$$
and in general with $x=r^2$
\begin{equation}
 P(r)={2r\over M}{(1-x)^{N-M-1}\over(N-M-1)!}
\left({d\over dx}\right)^{N-M}
{(1-x^N)\over 1-x}.
\label{G8}
\end{equation}

There are two important limiting cases for large $M$:
$\mu=M/N$ fixed and
$L=N-M$ fixed. For fixed $\mu$ and $M$ to $\infty$ we find the
mentioned
scaling behaviour:
\begin{equation}
 P(r)=  \bigl( {1 \over \mu }-1 \bigr) { 2r \over(1-r^2)^2}
\label{prmu}
\end{equation}
 for $r^2<{ \mu}$ and
$P(r)=0$ otherwise. The distribution shows a gap near the unit circle.
This gap resembles the one obtained for resonances in the chaotic scattering
problem for large number of channels \cite{HS92}. In this strongly non unitary
limit we are also able to simplify the cluster function

\begin{equation}
Y(z,z+\delta) = (M\rho (z))^2\ \exp (-\pi M\rho (z)\ |\delta |^2)
\label{cluster}
\end{equation}
which is just the Ginibre behaviour \cite{Ginib,mehta} with the distance  
$\delta$ rescaled by the local mean level distance $1/\sqrt{M \rho(z)}$ given  
by equ.(\ref{prmu}) through $\rho(z)=P(r)/2\pi r$. The same can be shown for  
the nearest neighbour distance 
distribution obtained by Grobe et al. \cite{Grobe} and  
applied to a damped chaotic kicked top.

In the other limit of fixed $L=N-M$ and $M$ to $\infty$, which may be
considered
as weakly nonunitary, we recover exactly the universal resonance-width
distribution  \cite{FS97} for perfect coupling to $L$ channels with
$y=N(1-r)$

\begin{equation}
 \rho(y)={y^{L-1}\over(L-1)!}\left({-d\over dy}\right)^{L}{1-
e^{-2y}\over 2y}.
\label{weak}
\end{equation}
Similarly the cluster function obtained in this limit can be shown to coincide
with the one obtained by Fyodorov and Khorushenko \cite{FK99} for chaotic  
scattering with a finite number of perfectly coupled channels.
The statistics (\ref{weak}) has also been found by Kottos and Smilansky
\cite{Kottos} for chaotic scattering on graphs and by Gl\"uck et al.  
\cite{Kol99a,Kol99} for a model of crystal electron 
in the presence of dc and ac fields.
In both of these works the S-matrix is reduced to the resolvent of a  
subunitary matrix as is investigated in the present paper.

\section{Submatrices of unitary symmetric  matrices}

For several applications one uses symmetric unitary matrices typical for
the circular orthogonal ensemble. This case is relevant if the physical
system possesses time reversal symmetry, or any generalized
anti-unitary symmetry \cite{H91}. Let $U$ be a random unitary matrix
typical of CUE. It is easy to prove that the symmetric matrix
$W:=UU^{T}$ is typical to COE \cite{mehta}. We shall thus define the
ensemble of truncated symmetric unitary matrices $W_{[N,M]}$.
In the definition of this ensemble the position of the submatrix is
crucial. We take the left upper part of the symmetric matrix $W$, thus
the truncated matrices $W_{[N,M]}$ are symmetric.

The distributions $P(t)$ and $P(r)$ for the symmetric matrices generated
out of COE matrices of size $N=16$ are shown in Fig.7.
Each plot contains data from $10^4$ symmetric random unitary matrices.
Note the differences between these figures and the corresponding data
for orthogonal and unitary matrices presented in figures 1 and 4.
If the truncation of the matrix $W$ is performed asymmetrically
(e.g. we take the lower left submatrix), the distribution $P(r)$ becomes
closer to this corresponding to the truncations of random unitary matrices
$U_{[N,M]}$.

In the asymptotic limit the properties of the
ensemble of the truncations of symmetric matrices $W_{[N,M]}$
depends on the same scaling parameter $\mu =M/N$.
Moreover, the distribution $P(r)$ becomes close to the corresponding one
for the unitary ensemble described by the distribution (\ref{prmu}).
Therefore we may conjecture that the distribution
$P_{\mu}(r)$, which
describes the distribution of moduli of eigenvalues of truncated matrices
in the limit of large $N$,  is universal and  does not
depend on the initial ensemble of random matrices, provided $M/N$ is fixed.
This corresponds to results for resonances in the limit of $L/N$ fixed and
$N$ to $\infty$ \cite{HS92}. In the contrary, there are differences to be
expected in the limit of weakly nonunitary matrices: $N-M$ fixed and $M$ to  
$\infty$ \cite{SFT99,FK99}.

\vskip -1.4cm
\begin{figure}
\hspace*{-1.6cm}
\vspace*{-1.1cm}
\epsfxsize=10.5cm
\epsfbox{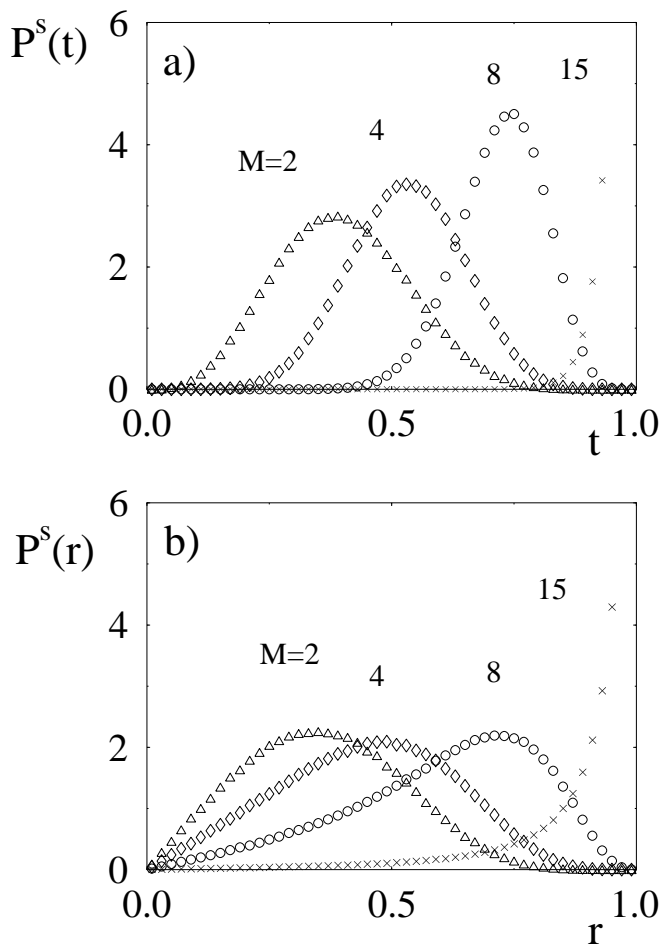}
\caption{Probability distributions $P(t)$ (a) and $P(r)$ (b)
for complex symmetric matrices $W_{[16,M]}$ with $M=2,4,8,15$. }
\label{fig7}
\end{figure}

\section{Concluding remarks }

Three families of ensembles of random matrices are proposed.
They are defined by cutting an $M$ dimensional submatrix of an initially
$N$--dimensional unitary matrix, pertaining to a given ensemble
of unitary, unitary symmetric or orthogonal matrices.
Using a link between the truncation of an orthogonal
matrix and the projection of a hypersphere into a smaller dimensional
space we found the probability distributions
of the lengths $t$ of the projected vectors.

Truncated matrices are not unitary and their complex
eigenvalues are located inside the unit circle.
We derived an analytical formula for the distribution
$P(r)$ of moduli of eigenvalues of truncations of the CUE matrices.
It takes a particularly simple form for small values of $N$ and $M$.
In the asymptotic limit $N\to \infty$ this distribution depends only
on the scaling parameter $\mu=M/N$, provided $M/N$ is not very close to $1$.
For small $r$ the distribution  $P_{\mu}(r)$ grows linearly,
later displays a nonlinear behaviour and eventually
suffers a sudden cut-off at $r_{\mu}=\sqrt{\mu}$.
For $N>>1$ the probability distribution $P(r)$
does not depend, whether the initial matrices are
orthogonal, unitary or unitary symmetric, again if $M/N$ is not very close
to $1$.
In this strongly nonunitary limit at least for the case of broken time reversal
symmetry correlations are shown to coincide with those obtained from the  
Ginibre ensemble of general complex matrices, provided distances are rescaled  
by the local mean level distance.
In the weakly nonunitary limit $N-M$ fixed and $M$ to $\infty$
the eigenvalue distribution
again for broken time reversal symmetry is shown
to coincide with the universal
resonance widths distribution in the weakly non-hermitian limit.
The same is true for the 2-point cluster function.

It is a pleasure to thank  F. Borgonovi, Y. Fyodorov,
G. Casati, R. G{\c e}barowski,
I. Guarneri, F. Haake, A. Kolovsky, M. Ku{\' s},
 and M. Nowak for inspiring discussions. One of us (K.{\.Z}) is grateful
to the International
Center for the Study of Dynamical Systems, Universita' degli Studi dell'
Insubria, where a part of this work was performed.
Financial support by Komitet Bada{\'n} Naukowych in Warsaw under
the grant 2P03B-00915 and by the Sonderforschungsbereich 'Unordnung
und gro{\ss}e Fluktuationen'  der Deutschen Forschungsgemeinschaft
is gratefuly acknowledged.

\appendix

\section{Convolution properties of the distributions
            $P(\lowercase{t})$ }

In the appendix we demonstrate the convolution properties of
the distributions $P^o_{N,M}(t)$ which might be used to derive the formula
(\ref{ponm}). We start thus with a random orthogonal matrix $O(N)$ or with
random points covering uniformly the hypersphere
$S^{N-1}$ of radius $1$. Their distribution in the polar coordinates is given by
Eq. ({\ref{domega}). For simplicity we will denote the distance from
the origin of a point projected into $R^M$ by $t_{N,M}$.
It is just the argument of the distribution
$P^o_{NM}(t)$. Due to the definition of the polar coordinates
$t_{N,N-1}=\sin\theta_{N-2}$,
$t_{N,N-2}=\sin\theta_{N-2} \sin\theta_{N-3}$, $\cdots$, and
$t_{N,1}=\cos\theta_{N-2}$. Therefore all variables $t_{N,M}$ 
 may be rewritten as the product consisting of $L=N-M$ factors
\begin{equation}
 t_{N,M}=  \prod_{k=1}^{N-M}  t_{N-k+1,N-k}.
\label{tprod}
\end{equation}
This factorization allows us to find the distributions (\ref{ponm}).

Probability distribution of a sum of two independent random variables
$z=x+y$ is given
by the
standard convolution $P_x\circ P_y:=P(z)= \int_{-\infty}^{\infty} P_x(x)
P_y(z-x)dx$.
In a similar way, the distribution of the product of two independent random  
variables,
$z=xy$,
is given by the product convolution
\begin{equation}
 P_x\star  P_y :=P(z) = \int_z^1 P_x(x) P_y
\bigl({z \over x}\bigr) { 1 \over  |x|} dx.
\label{prod}
\end{equation}
In the general case the integration should be performed over the entire
real axis, but in our case the intergration is restricted to $[z,1]$, since
all arguments $t\in [0,1]$.

Factorization (\ref{tprod}) allows us to write
convolution relations between probability distributions $P^o_{NM}(t)$.
For example
\begin{equation}
 P^o_{31}(t) = P^o_{32} \star  P^o_{21},
\label{pro1}
\end{equation}
\begin{equation}
 P^o_{42}(t) = P^o_{43} \star  P^o_{32},
\label{pro2}
 \end{equation}
\begin{equation}
 P^o_{41}(t) = P^o_{43} \star  P^o_{32} \star P^o_{21}.
 \label{pro3}
\end{equation}

 In general we obtain a convolution relation
\begin{equation}
 P^o_{N,M}(t) = P^o_{N,N-1} \star  P^o_{N-1,N-2}
              \star \cdots  \star  P^o_{M+1,M},
\label{pron}
\end{equation}
 which might be used to derive formula (\ref{ponm}).


\end{document}